# Primary User Emulation Attacks: A Detection Technique Based on Kalman Filter

**Zakaria El Mrabet[1*], Youness Arjoune[1], Hassan El Ghazi[2], and Badr Abou Al Majd[3], Naima Kaabouch[1]**

[1]  Department of Electrical Engineering, University of North Dakota, Grand Forks ND, USA;
    zakaria.elmrabet@und.edu; youness.arjoune@engr.und.edu; naima.kaabouch@engr.und.edu
[2]  STRS Lab, National Institute of Posts & Telecommunication, Rabat, Morocco; elghazi@inpt.ac.ma
[3]  Department of Mathematics, Mohammed V University, Rabat, Morocco; b.abouelmajd@fsac.ac.ma
*   Correspondence: zakaria.elmrabet@und.edu; Tel.: +1-701-885-2113

**Abstract:** Cognitive radio technology addresses the problem of the spectrum scarcity by allowing secondary users to use the vacant spectrum bands without causing interference to the primary users. However, several attacks could disturb the normal functioning of the cognitive radio network. Primary user emulation attack is one of the most severe attacks in which a malicious user emulates the primary user signal characteristics to either prevent other legitimate secondary users from accessing the idle channels or causing harmful interference to the primary users. There are several proposed approaches to detect the primary user emulation attackers. However, most of these techniques assume that the primary user location is fixed which does not make them valid when the primary user is mobile. In this paper, we propose a new approach based on Kalman filter framework for detecting the primary user emulation attacks with a mobile primary user. Several experiments have been conducted and the advantages of the proposed approach are demonstrated through the simulations.

**Keywords:** Cognitive radio; primary user emulation attacker; mobile primary user; Kalman filter; Received Signal Strength

## 1. Introduction

Cognitive radio (CR) technology is a viable solution that addresses the problem of the spectrum scarcity [1]. It enables secondary users to sense, dynamically adjust their transmission parameters, and access the idle frequency channels (spectrum holes) without causing any harmful interference to the primary users [2]. Due to the unreliable nature of the wireless communication, cognitive radio networks can be subject to various cyber-attacks which can have a negative impact on their performance [3-5]. Examples of these attacks include asynchronous sensing attacks [3], primary user emulation (PUE) attacks [5, 6], spectrum sensing data falsification (SSDF) attacks [6, 7], and jamming attacks [8, 9].

A PUE attack targets the CR physical and MAC layers and is considered as one of the most severe attacks in which a malicious user emulates the transmission characteristics of the primary user (PU) and mimics its behavior to mislead legitimate secondary users. Such an attack can create a harmful interference to the primary user and prevent other secondary users from using the idle spectrum frequency channels [5, 6]. There are two types of primary user emulation attackers [10]: selfish and malicious. The purpose of the selfish attacker is to use and selfishly exploit an idle frequency channel without sharing it with other legitimate secondary users. The malicious attacker, on the other hand, aims at causing a denial of service in cognitive radio networks, and preventing secondary users from accessing the available frequency channels.

Several approaches have been proposed to cope with the PUE attacks [11–24]. For instance, the authors of [11] proposed an energy-based detection approach to detect the source of the signal and decide if it is emitted by a legitimate primary user or an attacker. In their approach, each secondary user measures the power level of the received signal and compares it to that from a legitimate PU.

The authors of [12] proposed a belief propagation framework based on Markov random field to detect the primary user emulation attacker. Each secondary user decides whether the signal is coming from a legitimate primary user or not using the energy detection technique then calculates the belief and exchanges it with other secondary users. If the average of the belief values is lower than a predefined threshold, then the signal is coming from a malicious user, otherwise it is coming from a legitimate user. However, techniques based on energy detection are not efficient in distinguishing between noise and signal, and they suffer from a high probability of false alarm [13].

Feature-based techniques, such autocorrelation and matched filter [14, 15], are also inefficient in distinguishing between the PU signals and those of the PUE attacker. For instance, the authors of [16] used the cyclostationary feature of the transmitter's signal to detect the source of the incoming signal. However, this technique is not efficient in detecting malicious user which can mimic the primary user signal features. In [17], the authors proposed a radio-frequency fingerprinting detection technique. In this technique, the transmitter is identified based on some unique radiometric features extracted from its analog signals. In another paper [18], the authors proposed a detection technique using the characteristics of wireless channels. As the statistical property of the wireless channel between the transmitter and the receiver is unique in a wireless environment, this feature is used as a radio fingerprint to detect the primary user emulation attacker. However, radio fingerprinting based approaches require additional hardware or software to implement. In addition, these techniques are inefficient in identifying effectively the primary user signal since the characteristic of the noise introduced by the hardware is random.

Other techniques have been proposed to estimate the position of the transmitter and compare the estimated position with the known position of the primary user. The authors of [19] proposed a time difference of arrival (TDoA) based approach to estimate the position of the transmitter. In this approach, the time elapsed between the transmission of the signal and the reception of the reply is used to estimate the location of the transmitter. Thought TDoA can estimate more accurately the transmitter position than other techniques, it requires a tight synchronization between the transmitters and the receivers which is challenging. In [20], the authors proposed an angle of arrival based approach for detecting the transmitter's position. In this approach, the direction of the received signal is measured at different reference nodes then by applying the triangulation technique, the transmitter location can be estimated. However, this technique is affected by the multipath phenomenon [21]. The authors of [22] proposed a mitigation approach to distinguish the primary user signal from other signals via an energy-efficient localization technique and channel parameter variance. The authors of [23] proposed a model based on the trilateration, the received signal strength (RSS), and the particle swarm optimization to increase the detection accuracy of the primary user emulation attacker. All the previously mentioned techniques do not deal with uncertainty which affects the measurements. The authors of [24] proposed a Bayesian model and trilateration technique for detecting the primary user emulation attack position. Based on the received signal strength, the Bayesian decision theory is used to deal with the uncertainty related to the primary user environment and increase the detection accuracy of the primary user position.

The existing localization techniques assume that the primary user position is fixed and known. However, in wireless communication networks where the primary user is mobile, such as cognitive radio ad hoc networks, these techniques are inefficient. Therefore, in this paper, we propose a localization approach to detect the primary user emulation attacks with a mobile primary user. The proposed approach is based on the Kalman filter framework for predicting the location of a mobile primary user and detecting the primary user emulation attacker.

The remainder of this paper is organized as follows. In section II, we describe the proposed localization approach based on Kalman filter. In section III, we first present the metrics used for evaluating the proposed approach's performance, then we discuss some examples of results and compare the proposed approach with the RSS-based location technique. Finally, some conclusions are drawn in the last section.

## 2. Methodology

The proposed approach is based on the Kalman filter framework for tracking the primary user location and deciding if the incumbent signal is emitted by a legitimate primary user or from an attacker. Figure 1 shows the flowchart of the proposed approach. The primary user position is tracked using Kalman filter [25], then the distance, $d_{kfi}$, between a node $i$ and the legitimate primary user is estimated. Next, the received power of the transmitter is used to calculate the distance, $d_{pi}$, between the node $i$ and that transmitter using the Free Space Path Loss equation. If the difference between $d_{kfi}$ and $d_{pi}$ is greater or equal to a predefined threshold $\tau$, then the transmitter is an attacker. Otherwise, it is a legitimate primary user.

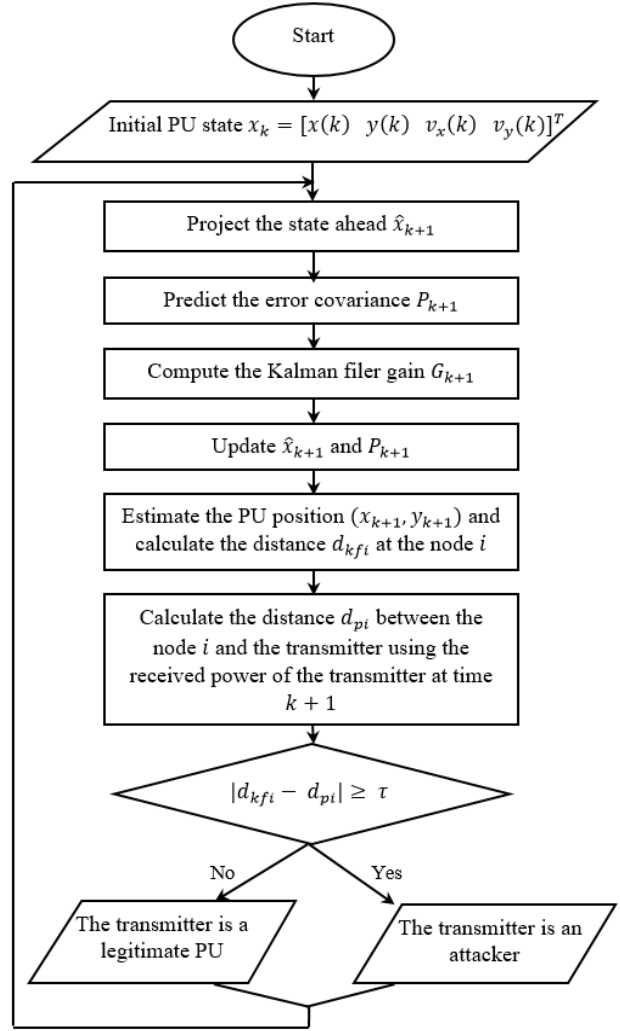

**Figure 1.** Flowchart of the proposed approach

In this work, we assume that the primary user is moving in a two-dimensional field and its state is described by its position and its velocity. This state is represented as:

$$x_k = [x(k) \quad y(k) \quad v_x(k) \quad v_y(k)]^T \tag{1}$$

Where $(x(k), y(k))$ are the coordinates of the primary user position at the time $t_k$ and $(v_x(k), v_y(k))$ are the velocities of the primary user in $x$ and $y$-directions at time $t_k$, respectively. In addition, we assume that the movement of the primary user is locally linear within the sampling interval. Thus, its motion can be modeled as:

$$x_{k+1} = A_k x_k + B_k u_k + w_k \tag{2}$$

Where $A$ and $B$ represent the transition matrix and the control matrix, respectively. $u_k$ is the acceleration of the primary user and $w_k$ is a white Gaussian noise with zero mean and covariance matrix $Q$, which is given by:

$$Q = \begin{bmatrix} \sigma^2_{wx} & 0 \\ 0 & \sigma^2_{wy} \end{bmatrix}$$

Where $\sigma^2_{wx}$ and $\sigma^2_{wy}$ are the covariances of $wx$ and $wy$ which correspond to the acceleration noise of the primary user along the X-axis and Y-axis, respectively. The transition and the control matrix are given by:

$$A_k = \begin{bmatrix} 1 & 0 & \Delta t_k & 0 \\ 0 & 1 & 0 & \Delta t_k \\ 0 & 0 & 1 & 0 \\ 0 & 0 & 0 & 1 \end{bmatrix}, \quad B_k = \begin{bmatrix} \frac{\Delta t k^2}{2} \\ \frac{\Delta t k^2}{2} \\ \Delta t_k \\ \Delta t_k \end{bmatrix}$$

Where $\Delta t_k = t_{k+1} - t_k$ is the sampling interval time between two successive measurements at $t_{k+1}$ and $t_k$. The measurement model $z_k$ adopted is given by:

$$z_k = C x_k + v_k \tag{3}$$

Where $C$ is the measurement matrix and $v_k$ is the measurement noise which is a white Gaussian noise with zero mean and covariance matrix $R$. $C$ is given by:

$$C = \begin{bmatrix} 1 & 0 & 0 & 0 \\ 0 & 1 & 0 & 0 \end{bmatrix}$$

During the first process of Kalman filter, the state of the primary user $\hat{x}_{k+1}$ and its associated covariance error matrix $P_{k+1}$ at time $t_{k+1}$ are predicted using the following equations:

$$\hat{x}_{k+1} = A_k \hat{x}_k + B_k u_k, \tag{4}$$

$$P_{k+1} = A_k P_k A_k^T + Q_k, \tag{5}$$

Where $\hat{x}_k$ is the primary user state at time $k$, $P_k$ is the covariance error matrix at time $k$, and $A_k^T$ is the transpose of transition matrix $A$ at time $k$. During the update process of Kalman filter, the estimated state and its covariance error matrix at time $k+1$ are updated and corrected using the Kalman filter gain $G_{k+1}$ as follows:

$$G_{k+1} = P_{k+1} C^T (C P_{k+1} C^T + R)^{-1}, \tag{6}$$

$$\hat{x}'_{k+1} = \hat{x}_{k+1} + G_{k+1}(z_k - C \hat{x}_{k+1}), \tag{7}$$

$$P'_{k+1} = P_{k+1} - C \hat{x}_{k+1} P_{k+1}, \tag{8}$$

Where $C^T$ is the transpose matrix of the measurement matrix $C$, $\hat{x}'_{k+1}$ is the updated state of the primary user state at $k+1$, and $P'_{k+1}$ is the updated covariance error matrix $P_{k+1}$. Once the coordinates of the primary user position $(x_p, y_p)$ are estimated using Kalman filter, the distance $d_{kfi}$ between a fixed position of an anchor node $(x_i, y_i)$ and the primary user can be obtained using the following equation:

$$d_{kfi} = \sqrt{(x_p - x_i)^2 + (y_p - y_i)^2}, \tag{9}$$

In order to verify if the incoming signal is emitted by a legitimate primary user or by an attacker, a distance $d_{pi}$ between the transmitter and a node $i$ is required. This distance can be obtained from the received power signal of the transmitter using the Free Space Path Loss equation [24]:

$$P_r = \frac{P_t G_t G_r \lambda^2}{(4\pi d_{pi})^2} \tag{10}$$

Where $P_r$ is the received power, $P_t$ is the transmitted power, $G_r$ is the antenna gain of the receiver, $G_t$ is the antenna gain of the transmitter, $\lambda$ is the wavelength, and $d_{pi}$ is the distance between an anchor node $i$ and the transmitter. The equation (10) can be expressed in dB as:

$$P_r(dB) = -10\alpha \log_{10}(d_{pi}) + A \tag{11}$$

Where $\alpha$ is the propagation path loss exponent and $A$ is expressed as:

$$A = 10log_{10}(P_t G_t G_r \lambda^2) - 20log_{10}(4\pi) \qquad (12)$$

The received signal strength at an anchor node $i$ can be impacted by the noise, the equation (12) can be written as:

$$P_r(dB) = -10\alpha log_{10}(d_{pi}) + A + n_i \qquad (13)$$

Where $n_i$ is a white Gaussian noise that follows the normal distribution $N(0, \sigma^2)$. Thus, the distance $d_{pi}$ can be estimated as:

$$d_{pi} = 10^{\frac{(A+n_i-pr)}{10\alpha}} \qquad (14)$$

By comparing the difference between the estimated distance $d_{kfi}$ and $d_{pi}$ at an anchor node $i$ to a predefined threshold $\tau$, the transmitter is considered as a primary user emulation attacker if the $|d_{kfi} - d_{pi}| \geq \tau$ and as a legitimate primary user when the $|d_{kfi} - d_{pi}| < \tau$.

## 3. Results

Due to a number of random variables used in the proposed approach, including the acceleration noise, the measurement noise and the signal noise, the Monte Carlo simulation was used to handle the uncertainty in the simulation. The proposed approach was implemented in Matlab and extensively tested and evaluated using several metrics including the probability of detection, the probability of false alarm, and the probability of miss detection. Additionally, the proposed approach was compared to an RSS-based localization technique.

The probability of detection, $P_d$, corresponds to the number of times where the primary user emulation attacker signal is correctly classified as an attack divided by the total number of trials. It is given by:

$$P_d = \frac{Number\ of\ PUE\ detections}{Number\ of\ total\ trials} \qquad (15)$$

The probability of false alarm, $P_{fa}$, corresponds to the number of times where the legitimate primary user signal is wrongly classified as an attack divided by the total number of trials. It is expressed as:

$$P_{fa} = \frac{Number\ of\ false\ detected\ attack}{Number\ of\ total\ trials} \qquad (16)$$

The probability of miss detection, $P_m$, corresponds to the number of times where an attack is incorrectly classified as a normal signal divided by the total number of trials. It is given by:

$$P_m = \frac{Number\ of\ miss\ detections}{Number\ of\ total\ trials} \qquad (17)$$

Examples of results are given in Figure. 2 through 7. Figure 2 shows the scenario used for tracking the primary user trajectory with Kalman filter. As one can see, the predicted trajectory is almost the same as the actual trajectory of the legitimate primary user.

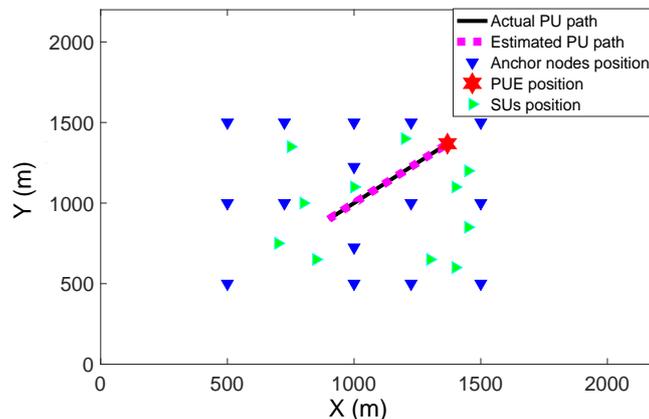

**Figure. 2.** Scenario used for tracking the primary user with Kalman filter.

Figure 3 shows the probability of detection as a function of the distance, $d_{pu\_pue}$, between a mobile primary user and a primary user emulation attacker for different SNR values. As one can see, the probability of detection increases as the distance between the primary user and the attacker increases. For example, for an SNR value of -10dB, the probability of detection is equal to 37% when the distance between the primary user and the attacker is 50m, and this probability increases to 60% when the distance is equal to 100m. In addition, this figure shows that the probability of detection increases with the increase of SNR values. For instance, for a distance of 50m between the PU and the PUE attacker, the probability of detection is equal to 37%, 80%, 96%, 99,2%, 99,7% for an SNR value corresponding to -10dB, -5dB, 0 dB, 5dB, 10dB, respectively.

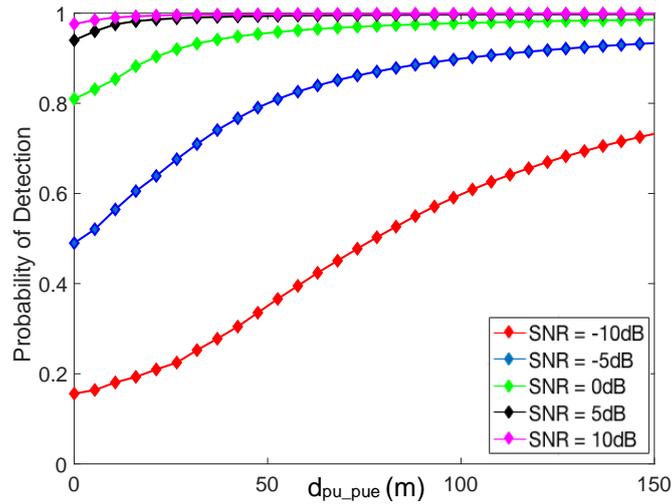

**Figure. 3.** The probability of detection as a function of the distance $d_{PU\_PUE}$ between a mobile PU and a PUE.

Figure. 4 shows the probability of detection as a function of the probability of false alarm for a distance $d_{pu\_pue}$ of 30m between the primary user and the primary user emulation attacker. As it can be observed, the probability of detection increases with the increase of the probability of false alarm. For example, for an SNR value of -15dB and with a probability of false alarm of 10%, the corresponding probability of detection is 16%, and when the probability of false alarm increases to 30%, the probability of detection is equal to 61%. In addition, with the increase of the SNR value, the probability of detection increases. For instance, with a probability of false alarm of 2%, the probability of detection is equal to 39%, 63% 90%, 99%, and 100% for an SNR value corresponding to -15dB, -10dB, -5dB, 0dB, and 5dB, respectively.

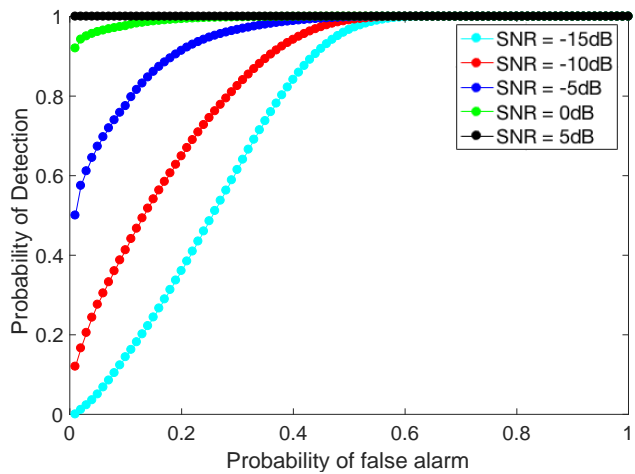

**Figure. 4.** Probability of detection as a function of the probability of false alarm for $d_{PU\_PUE} = 30$m.

Figure. 5 shows the probability of miss detection as a function of the distance $d_{pu\_pue}$ between a mobile primary user and a primary user emulation attacker. As it can be observed, the probability of miss detection decreases as the distance between the PU and the PUE increases. For example, for an SNR value of -10dB, the probability of miss detection is equal to 65% when the distance between the primary user and the attacker is equal to 50m, and it decreases to 41% when the distance increases to 100m. In addition, this figure shows that when the SNR value increases, the probability of miss detection decreases. For example, with a distance of 110m between the primary user and the attacker, the probability of miss detection is equal to 37%, 10%, 5%, 1%, and 0% for SNR values of -10dB, -5dB, 0dB, 5dB, 10dB, respectively.

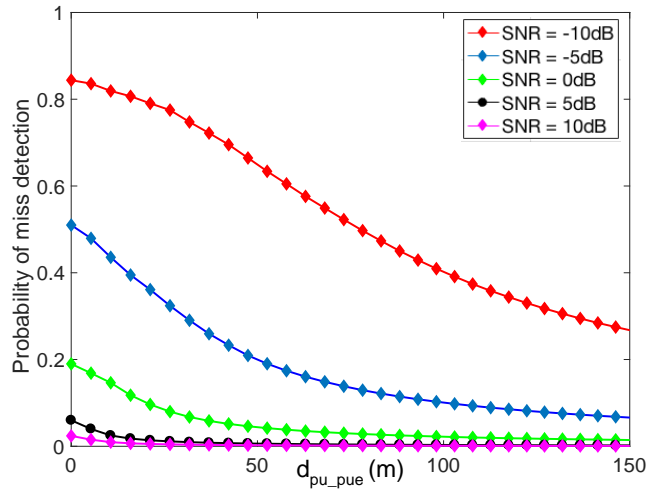

**Figure. 5.** Probability of miss detection as a function of the distance $d_{PU\_PUE}$ between the mobile PU and a PUE.

Figure. 6 shows a comparison between the proposed approach and an RSS-based localization technique in terms of the probability of detection as a function of distance with an SNR value of -10dB. As shown in this figure, the probability of detection for the proposed as well as the RSS-based localization approaches is low when the attacker is in a close proximity to the legitimate primary user location. When the distance increases, the probability of detection of the proposed approach increases while it remains almost the same for the RSS-based localization technique. For instance, when the distance between the primary user emulation attacker and the primary user is equal to 50m, the probabilities of detection of the proposed approach and the RSS-based localization technique are 24% and 16%, respectively. When the distance increases to 100m, the probability of detection of the proposed approach increases to 49% while it remains equal to 16% for the RSS-based localization approach.

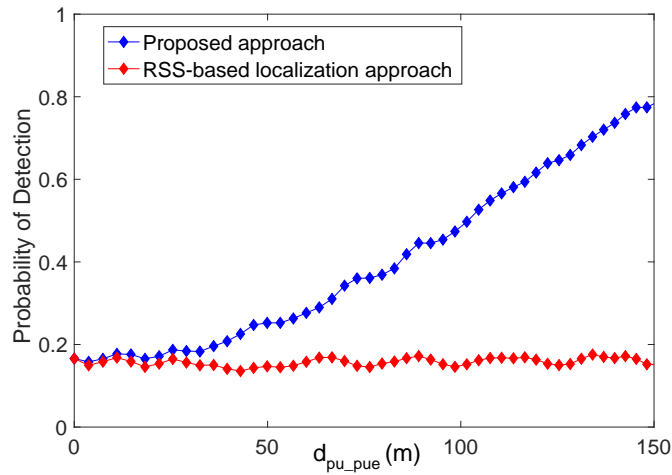

**Figure. 6.** Comparison between the proposed technique and the RSS-based localization technique in terms of the probability of detection as a function of the distance.

Figure. 7 illustrates a comparison between the proposed technique and the RSS-based localization technique in terms of the probability of miss detection as a function of the distance with an SNR value equal to -10dB. As one can see, the probabilities of miss detection of the proposed approach and the RSS-based localization technique are high when the attacker is close to the primary user location. When the primary user starts moving and the distance becomes larger, the probability of miss detection of the proposed approach decreases while it remains the same for the RSS-based localization technique. For example, when the distance is 50m, the probabilities of miss detection of the proposed approach and the RSS-based localization technique are 76% and 84%, respectively. When the distance increases to 150m, the probability of miss detection of the proposed approach decreases to 19% while it remains equal to 84% for the RSS-based localization technique.

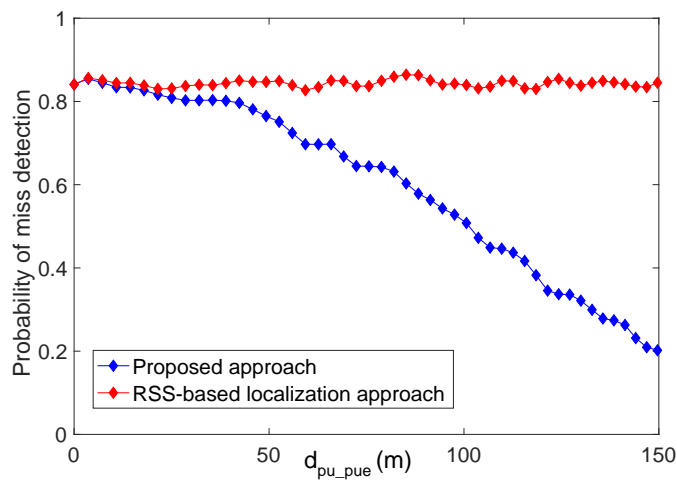

**Figure. 7.** Comparison between the proposed technique and the RSS-based localization technique in terms of the probability of miss detection as a function of the distance.

## 4. Conclusions

Cognitive Radio networks are subjects to several cyber-attacks. Primary user emulation attack is one of the most severe attacks that can impact the normal functioning of these networks. In this paper, we propose a new approach for detecting the primary user emulation attacker with a non-stationary primary user. Kalman filter is used for tracking and estimating the position of the mobile primary user, then the received power of the transmitter is used to detect any potential primary user emulation attacker. Several experiments have been conducted and the model has been extensively tested and compared to the RSS-based location approach. The results show that the proposed approach produces satisfactory results in terms of tracking the primary user in a non-stationary environment and it outperforms the RSS-based localization technique. However, the proposed technique has a few limitations that need to be addressed in future works. These limitations include finding the initial coordinates of the primary user, handling the uncertainty in measurements, and dealing with non-linear system dynamics.